\documentstyle[12pt]{article}
\begin{document}
\title{ENERGETICS OF THE  EINSTEIN--ROSEN SPACETIME}
\author{L. Herrera$^{1}$\thanks{e-mail: laherrera@cantv.net.ve}, A. Di Prisco$^{1}$\thanks{e-mail: adiprisc@fisica.ciens.ucv.ve},   J.
Carot$^{2}$\thanks{e-mail: jcarot@uib.es} and N. O.
Santos$^{3,4,5}$\thanks
{e-mail: N.O.Santos@qmul.ac.uk and santos@ccr.jussieu.fr}\\
\small{$^1$Escuela de F\'{\i}sica, Facultad de Ciencias,} \\
\small{Universidad Central de Venezuela, Caracas, Venezuela.}\\
\small{$^2$Departament de  F\'{\i}sica,}\\
\small{Universitat Illes Balears, E-07122 Palma de Mallorca, Spain}\\
\small{$^3$School of Mathematical Sciences, Queen Mary, University
of London,}\\
\small{London, UK.}\\
\small{$^4$Laborat\'orio Nacional de Computa\c{c}\~{a}o Cient\'{\i}fica,}\\
\small{25651-070 Petr\'opolis RJ, Brazil.}\\
\small{$^5$LERMA/CNRS-FRE 2460, Universit\'e Pierre et Marie Curie, ERGA,}\\
\small{Bo\^{\i}te 142, 4 Place Jussieu, 75005 Paris Cedex 05,
France.}} \maketitle

\vspace{-0.5cm}

\begin{abstract}
A study covering some aspects of the Einstein--Rosen metric is presented. The electric and magnetic parts of the Weyl tensor are
calculated. It is shown that there are no purely magnetic E--R spacetimes, and also  that a purely electric E--R spacetime is  necessarily static. The
geodesics equations  are found and circular ones are analyzed in detail. The super--Poynting and the ``Lagrangian'' Poynting vectors are calculated and their
 expressions are found for two specific examples. It is shown that for a pulse--type solution, both expressions describe an inward radially directed flow of energy, far behind the wave
front. The physical significance of such an effect is discussed.
\end{abstract}
\newpage
\date{}

\section{Introduction}
Einstein--Rosen (E--R) spacetime \cite{ER} has attracted the
attention of researchers for many years (see
\cite{Thorne}--\cite{Herrera} and references therein). Its interest
may be understood by recalling that E--R represents the simplest
example of a spacetime describing outgoing gravitational waves.

It is our purpose in this work to study some aspects of the E--R
spacetime which have not been considered until now, and which may
shed some light on the nature of such a physically meaningful
spacetime.

Particular attention  deserves the question of whether or not
cylindrical gravitational waves have energy and momentum. Indeed,
this problem has been investigated by many authors (see
\cite{Rosen1}-\cite{Vir} and references therein) using different
energy--momentum pseudotensors. However such a practice although widely accepted (e.g. see \cite{vir} and references therein ) has been
questioned by many physicists, for the obvious reason that those
objects are not tensors (see \cite{Valeria}). Fortunately though, we
have available a tensor quantity which allows to define a covariant
(super)--energy density and a (super)--Poynting vector, namely the
Bel--Robinson tensor \cite{Bel}. We shall study the energetics of
E--R spacetime with the help of these quantities.

With this goal in mind, we shall first calculate the electric
$E_{\alpha \beta}$ and magnetic $H_{\alpha \beta}$ parts of Weyl
tensor, as well as the two invariants obtained from them, for the
Einstein--Rosen metric. These geometric objects have been under
discussion  for many years (see \cite{Bel}--\cite{Ferrando} and
references therein). Not only because of the  eventual relationship
of the magnetic part of the Weyl tensor with rotation \cite{glass,
Bonnor}, but also because of its link  with gravitational radiation
\cite{Bel, Bruni, Bonnora, Dunsby, Maartens, Hogan, HerreraII}.

From the expressions obtained for the above objects, it  follows (although the presented proof is somehow restricted)
that the vanishing of the magnetic part implies that the spacetime
is static, whereas the vanishing of the electric part of the Weyl
tensor is shown to imply that the spacetime is Minkowski, in
agreement with the fact that purely magnetic vacuum
space--times do no exist \cite{Wade, Hadow, Bonnor, Bergh,
Ferrando}.

Next we obtain the geodesic equations for a test particle in an E--R
spacetime. We analyze the circular motion and compare the behaviour
of the test particle in the static case with respect to its
behaviour in two (non--static) solutions of the E--R family, namely:
a pulse--type solution (\cite{Carmeli}, \cite{Herrera}) and the
Kramer solution \cite {Kramer} (both in a static background). 

The motivation for doing so become clear when  the
resulting differences (with respect to the static situation) are
analyzed and interpreted with the help  of the expression of the
radial component of the super--Poynting vector \cite{Basset}, and the Poynting
(pseudo)--vector defined by Stephani \cite{Stephani}. 

In the first   example (pulse--type solution) the angular  velocity (far behind the front)  is smaller than in the static case, this
difference decreases asymptotically as the system gets back to the
static situation. The physical interpretation of this result will be given in terms of the energetics of the system (see the last section).

In the case of the Kramer solution (in a static background), the time average of the  tangential
velocity of the particle is greater than in the static case. This suggests  an increasing of the
gravitational mass of the source, which might be related to an
inward radially directed flow of energy. However,  the time average
of the Stephani vector vanishes, and,  unfortunately, it was
impossible to evaluate the time average of the super--Poynting
vector, even with the help of a computer, although it seems that
such an average is different from zero.  An alternative
interpretation, not involving an inward flux of energy, is also
presented.

At this point it is worth mentioning that  in a  general time--dependent E--R spacetime, specific constraints have to be satisfied, for   circular geodesics to exist (see
secion 5.1 below). Accordingly, in the two examples examined here, we shall define the conditions under which, circular geodesics do in fact exist. Of course all the discussion will be
carried on, upon satisfaction of the above mentioned conditions. Also it is worth stressing that both solutions considered here are defined in a static background.

Thus, in the case of the pulse (in a static background) we shall see that conditions for the existence of circular orbits are satisfied, behind the pulse,  very far from the front
(in 5.1 we specify what we mean by ``very far'') where  the energy flux (as defined by the super--Poynting or the Stephani vector) is negative (inward).

 It is worth noticing that this last result is not in contradiction with the fact that the rate of change of $C$--energy is shown to be negative in \cite{Apostolatos}, since that
proof is only valid for large values of $r$, where in fact the radial component of, both, the super--Poynting and the Stephani vectors, are positive (see section 6).

In the case of the Kramer spacetime (in a static background), conditions for the existence of circular geodesics are satisfied only for the time average of the metric.

\section{The Einstein--Rosen metric and its electric and magnetic Weyl tensor}
The line element reads
\begin{equation}
ds^2=-e^{2\gamma-2\psi}(dt^2-dr^2)+e^{2\psi}dz^2+r^2 e^{-2\psi}d\phi^2
\label{1}
\end{equation}
where $\psi=\psi(t,r)$ and $\gamma=\gamma(t,r)$ satisfy the Einstein equations:
\begin{equation}
\psi_{tt}-\psi_{rr}-\frac{\psi_r}{r}=0
\label{2}
\end{equation}
\begin{equation}
\gamma_{t}=2r\psi_{r}\psi_t
\label{3}
\end{equation}
\begin{equation}
\gamma_{r}=r(\psi_{r}^2+\psi_t^2)
\label{4}
\end{equation}
where indexes stand for differentiation with respect to $t$ and $r$.

Now, for an observer at rest in the frame of (\ref{1}), the four-velocity
vector has components
\begin{equation}
u^{\alpha} = \left(e^{\psi-\gamma}, 0, 0, 0\right)
\label{5}
\end{equation}
and it is obvious that for such a congruence of observers the
vorticity $\omega_{\alpha \beta}$ vanishes.

\subsection{The electric and magnetic  parts of Weyl tensor}
The electric and magnetic parts of Weyl tensor, $E_{\alpha \beta}$ and
$H_{\alpha\beta}$, respectively, are formed from the Weyl tensor $C_{\alpha
\beta \gamma \delta}$ and its dual
$\tilde C_{\alpha \beta \gamma \delta}$ by contraction with the four--velocity vector given by (\ref{5}):
\begin{equation}
E_{\alpha \beta}=C_{\alpha \gamma \beta \delta}u^{\gamma}u^{\delta}
\label{electric}
\end{equation}
\begin{equation}
H_{\alpha \beta}=\tilde C_{\alpha \gamma \beta \delta}u^{\gamma}u^{\delta}=
\frac{1}{2}\epsilon_{\alpha \gamma \epsilon \delta} C^{\epsilon
\delta}_{\quad \beta \rho} u^{\gamma}
u^{\rho},
\qquad \epsilon_{\alpha \beta \gamma \delta} \equiv \sqrt{-g}
\;\;\eta_{\alpha \beta \gamma \delta}
\label{magnetic}
\end{equation}
where
$\eta_{\alpha\beta\gamma\delta}= +1$ for $\alpha, \beta, \gamma, \delta$ in
even order, $-1$ for $\alpha, \beta, \gamma, \delta$ in
odd order and $0$ otherwise.

We have calculated the magnetic and electric part using GRTensor.
Thus, one has for the components of the magnetic Weyl tensor:
\begin{equation}
H_{23}=H_{32} = r e^{2(\psi -\gamma)}\psi_t \left[ -3 \psi_r -
\frac{\psi_{t r}}{\psi_t} + r \left(\psi_t^2  + 3 \psi_r^2 \right)
\right]  \label{h23} \end{equation} and the rest of the components being
zero.

Regarding the electric part, one gets:

\begin{equation} E_{11} = \psi_t^2 -\psi_r^2 + \frac{1}{r} \psi_r  \label{e11} \end{equation}
\begin{equation}
E_{22} =  e^{ (4 \psi -2 \gamma)}\left[- 2\psi_t^2  - \psi_r^2
-\psi_{r r} - \frac{\psi_r}{r} + r\psi_r \left( 3\psi_t^2 +
\psi_r^2\right)\right] \label{e22} \end{equation}
\begin{equation}
E_{33} = r^2 e^{-2\gamma} \left[\psi_{r r} - r\psi_r \left(
3\psi_t^2 + \psi_r^2 \right) + \psi_t^2 + 2\psi_r^2\right] \label{e33} \end{equation}\\

Notice, however, that these components  are not all independent,
since they satisfy the relation
\begin{equation}
E_{22}e^{2\gamma-4\psi}+\frac{E_{33}}{r^2}e^{2\gamma}=-E_{11}
\label{rel}
\end{equation}

The two invariants

$$ Q = H^{a}_{\; b} E^{b}_{\; a}, \qquad I = E^{a}_{\; b} E^{b}_{\;
a} - H^{a}_{\; b} H^{b}_{\; a}$$ can now be easily computed to give:

\begin{equation} Q=0\end{equation} \label{Q}

\begin{eqnarray}
I&=&-\frac{2}{r^2}e^{4(\psi-\gamma)} \left\{(\psi_t -
\psi_r)^3 (\psi_t +\psi_r)^3 r^4   \right.\nonumber\\
&& + r^3 \left[ -2\psi_t\psi_{tr}(3\psi^2_r +\psi_t^2) + 3\psi_r^5 +
(2\psi_{rr} - 6\psi_t^2)\psi_r^3 + 3\psi_r\psi_t^2(\psi_t^2 +
2\psi_{rr}) \right] \nonumber\\
&& + r^2\left[\psi_{tr}^2 + 6 \psi_{tr} \psi_t\psi_r - 2 \psi_r^4 +
9(\psi_t^2 - 3\psi_{rr})\psi_r^2 -\psi_{rr}^2 -
3\psi_t^2(\psi_t^2+\psi_{rr}) \right]\nonumber\\
&&\left. -r\psi_r(3\psi_t^2+\psi_{rr}) -\psi_r^2\right\}
\label{L}
\end{eqnarray}

\section{Purely magnetic solutions (PMS)}
Let us now see that  there exist no purely magnetic solutions, i.e. solutions satisfying $E_{\alpha \beta}=0$, in agreement with the known fact that purely magnetic vacua with
a non--rotating congruence are flat indeed.

From (\ref{e11})-(\ref{e33}) one obtains

\begin{equation}
6\psi_{r}^2+\psi_{rr}-\frac{\psi_{r}}{r}-4r \psi_{r}^3=0
\label{7}
\end{equation}
whose first general integral is
\begin{equation}
\psi_{r}=\frac{1}{2r} \pm \frac{1}{2r\sqrt{1-4r^2c}}
\label{8}
\end{equation}
with $c$  an arbitrary function of time, but since the range of $r$
extends to infinity, it turns out that $c=0$.

Thus, either $\psi_r=0$ which implies Minkowski, or
\begin{equation}
\psi=\ln r+\beta
\label{9}
\end{equation}
where $\beta$ is an arbitrary function of time.

However, from $E_{11}=0$ and (\ref{9}) it follows that $\beta$ is a
constant, then implying that $\psi_t=0$ and hence $H_{\alpha
\beta}=0$, which produces a Minkowski spacetime. Accordingly, we
conclude that, as expected, there are no PMS Einstein-Rosen waves.
\section{Purely electric solutions (PES)}
Let us now see if there exist solutions satisfying the condition $H_{\alpha \beta}=0$.

It is clear  from (\ref{h23})  that $\psi_t=0$ satisfies the
condition  $H_{\alpha \beta}=0$. The question is whether or not
there exist solutions with $\psi_t \neq 0$ also satisfying
$H_{\alpha \beta}=0$. We shall see that this is not the case.

Indeed, the general form of $\psi$, satisfying the wave equation (\ref{2}) (see \cite{Arfken}, \cite{Thorne}), with the outgoing wave condition (i.e. no inward travelling waves), is
\begin{equation}
\psi=Re \int_{0}^{\infty}A(\omega)e^{-i\omega t}H^{(1)}_{0} (\omega r)d\omega
\label{10}
\end{equation}
where $H^{(1)}_{0} (\omega r)$ is the Hankel function of the first
kind. Then it follows

\begin{equation}
\psi_t=Re \int_{0}^{\infty}-i\omega A(\omega)e^{-i\omega t}H^{(1)}_{0} (\omega r)d\omega
\label{11}
\end{equation}
\begin{equation}
\psi_r=Re \int_{0}^{\infty}-\omega A(\omega)e^{-i\omega
t}H^{(1)}_{1} (\omega r)d\omega \label{12}
\end{equation}
and
\begin{equation}
\psi_{tr}=Re \int_{0}^{\infty}i\omega^2 A(\omega)e^{i\omega
t}H^{(1)}_{1} (\omega r)d\omega \label{13}
\end{equation}
where it has been used the fact that \cite{Lebedev}
\begin{equation}
\frac{d}{dz}[z^{-\nu}H^{(p)}_{\nu}(z)]=-z^{-\nu}H^{(p)}_{\nu+1}(z)
\label{14}
\end{equation}

Next, let us consider the low frequency regime ($\omega
\stackrel{<}\sim \frac{1}{r}$). From the fact that $\psi$ is always
finite and $\int_{0}^{\infty}x^n H^{(1)}_{0}(x)dx$ converges only
for $n > -1$, it follows that as $\omega \rightarrow 0$,
$A(\omega)\sim \omega ^n$ with $n>-1$. This implies that the leading
terms in low-frequency ($\omega \stackrel{<}\sim \frac{1}{r}$)
contributions to $\psi_t$, $\psi_r$ and $\psi_{tr}$ are
\begin{equation}
\psi_t(\omega \stackrel{<}\sim\frac{1}{r}) \sim \psi_r(\omega
\stackrel{<}\sim \frac{1}{r}) \sim \frac{1}{r^{2+n}} \label{13b}
\end{equation}
and
\begin{equation}
\psi_{tr}(\omega \stackrel{<}\sim\frac{1}{r}) \sim \frac{1}{r^{3+n}}
\label{14b}
\end{equation}

From the above it is clear that the leading contribution in
(\ref{h23}), in the low frequency regime, comes from the term
$\psi_{tr}$, therefore the condition $H_{\alpha \beta}=0$ would
imply $\psi_{tr}=0$. Since this is clearly incompatible with the
general solution (\ref{10}), we have to assume $\psi_{t}=0$. 

Even though  the consistency of any solution must be assured for any value of $\omega r$ (including the low
frecuency regime),  it may be argued however that the proof is limited to the low frecuency regime and therefore is not completely general. 

So, let us present an alternative argument. Let us  supposse  that $A(\omega)$ has a compact support, i.e. $A=0$ for $\omega>\omega_c$, where
$\omega_c$ is some constant. Then close to the symmetry axis where $\omega r\ll 1$ we have (see \cite{Arfken},\cite{Lebedev})
\begin{equation}
H^{(1)}_{0}(\omega r)\approx-\frac{2 i}{\pi}\ln{\frac{2}{\omega r}}
\label{1refe}
\end{equation}

and

\begin{equation}
H^{(1)}_{1}(\omega r)\approx-\frac{2 i}{\pi}\frac{\Gamma(1)}{\omega r}
\label{2refe}
\end{equation}

Feeding back (\ref{1refe}) and (\ref{2refe}) into (\ref{h23}) it follows at once that condition $H_{\alpha \beta}=0$ implies  $\psi_{t}=0$. 
 
Thus, we may say 
that all PES are necessarily static. In agreement with the well known
fact that all static solutions are purely electric.

\section{The geodesics}
The geodesic equations may be obtained from the Lagrangian
\begin{equation}
{\cal L} =\frac{1}{2}g_{\mu \nu}\frac{d x^\mu}{ds}\frac{dx^\nu}{ds}
\label{geo1}
\end{equation}
then the Euler--Lagrange equations
\begin{equation}
\frac{d}{ds}\left(\frac{\partial {\cal L}}{\partial \dot
x^\alpha}\right)-\frac{\partial {\cal L}}{\partial x^\alpha}=0
\label{geo2}
\end{equation}
for the metric (\ref{1}) (dots denote derivatives with respect to
$s$)  become
\begin{eqnarray}
&&e^{2(\gamma - \psi)}\left[(\gamma_t-\psi_t)(\dot t^2+\dot r^2)+2(\gamma_r-\psi_r)\dot t \dot r +\ddot t \right] \nonumber \\
&+&e^{2\psi}\psi_t \dot z^2-e^{-2\psi}r^2 \psi_t\dot \phi^2=0
\label{geo3}
\end{eqnarray}
\begin{eqnarray}
&&e^{2(\gamma - \psi)}\left[2\dot t \dot r(\gamma_t-\psi_t)+(\dot t^2+\dot r^2)(\gamma_r-\psi_r) +\ddot r \right]\nonumber \\
&-&e^{2\psi}\psi_r \dot z^2+e^{-2\psi}(r^2 \psi_r-r)\dot \phi^2=0
\label{geo4}
\end{eqnarray}
\begin{equation}
\ddot z +2\dot z \dot t \psi_t+2\dot z \dot r \psi_r=0
\label{geo5}
\end{equation}
\begin{equation}
\ddot \phi +2 \dot \phi\frac{\dot r}{r} -2\dot r \dot \phi \psi_r-2\dot t \dot \phi \psi_t=0
\label{geo6}
\end{equation}
furthermore from (\ref{1}) we obtain
\begin{equation}
-\epsilon=-e^{2\gamma -2\psi}(\dot t^2-\dot r^2)+e^{2\psi}\dot z^2+r^2 e^{-2\psi}\dot \phi^2
\label{geo7}
\end{equation}
where $\epsilon=0,1$ or $-1$ if the geodesics are, respectively, null, timelike, or space--like.

Observe that (\ref{geo6})  may be writte as
\begin{equation}
\frac{d}{ds}(r^2e^{-2\psi}\frac{d\phi}{ds})=0,
\label{L1}
\end{equation}
implying that $L=r^2e^{-2\psi}\frac{d\phi}{ds}$ is a constant of motion.

\subsection{Circular geodesics}
Now we restrict ourselves to the case of circular geodesics, implying
\begin{equation}
\dot r=\dot z=0
\label{geo8}
\end{equation}
Then using (\ref{geo8}) in (\ref{geo3})--(\ref{geo7}) we obtain for the angular velocity
\begin{equation}
\omega=\frac{\dot \phi}{\dot t}
\label{geo9I}
\end{equation}
\begin{equation}
\omega^2=-\frac{e^{2\gamma}(\gamma_r-\psi_r)}{r^2\psi_r-r}
\label{geo9}
\end{equation}
or using the field equation (\ref{4})
\begin{equation}
\omega^2=\frac{e^{2\gamma}\psi_t^2}{1-r\psi_r}-e^{2\gamma}\frac{\psi_r}{r}
\label{geo10}
\end{equation}
From the definition for tangential velocity

\begin{equation}
V^\mu = (-g_{00})^{-1/2} U^{\mu} \label{defv}
\end{equation}
where
\begin{equation}
U^\mu = \left(\delta^\mu_\alpha + u^\mu u_\alpha\right) \frac{d
x^\alpha}{d t} \label{um}
\end{equation}
and $u^{\mu}$ is given by (\ref{5}), we obtain
\begin{equation}
V^\mu = e^{-(\gamma-\psi)}(0,0,0,\omega)
\label{vmc}
\end{equation}
thus
\begin{equation}
V^\mu V_\mu = e^{-2\gamma} r^2 \omega^2
\label{vv}
\end{equation}
which, using (\ref{geo10}) can be written as
\begin{equation}
V^\mu V_\mu = \frac{(r \psi_t)^2}{(1-r \psi_r)} - r \psi_r
\label{vv1}
\end{equation}

Next,  substituting (\ref{L1}) into (\ref{geo7}) and using (\ref{geo5}) we obtain for circular orbits 
\begin{equation}
L^2=-\epsilon r^3e^{-2\psi}\times
\frac{r\psi_{,t}{}^2+r\psi_{,r}{}^2-\psi_{,r}}
{r^2\psi_{,t}{}^2+r^2\psi_{,r}{}^2-1}.
\end{equation}

The above equation is a necessary condition for the existence of the 
circular orbit. Since the left hand side of the above equation is 
constant, the right hand side should be independent on the time $t$ too. This specific constraint should be satisfied in order to assure the existence of circular geodesics in any E--R
spacetime. 

This is indeed the case, for the two examples examined here. Thus, in the case of the pulse (in a static background), very far from the front ($t>>r$) we shall obtain (see below) 
\begin{equation}
L^2 \approx L_{LC}^2+ O(\frac{r}{t})
\label{L2}
\end{equation}
where $L_{LC}$ correspond to the static (Levi--Civita) solution. Thus, whenever we can neglect terms of the order $O(\frac{r}{t})$ and higher, we can safely assume that circular
gedodesics do exist.

In the case of the Kramer spacetime (in a static background), we shall consider only the time average of the corresponding variables, in which case is obvious that the requirement
above on $L$ is also satisfied.
\subsection{Static case}
This case is represented by the well known Levi--Civita solution \cite{Levi}, \cite{Mac}, \cite{HSG}
\begin{equation}
\psi_{LC} = \alpha - \beta \ln{r} \quad \alpha,\beta \; {\rm constants}
\label{psist}
\end{equation}
and
\begin{equation}
\gamma_{LC} =\beta^2 \ln{r}
\label{gamst}
\end{equation}
Using (\ref{psist}) into (\ref{geo10}) and (\ref{vv1})
we obtain $\omega^2$ and $V^\mu V_\mu$ for this case
\begin{equation}
\omega^2 = \beta r^{2(\beta^2-1)}
\label{w2st}
\end{equation}
\begin{equation}
V^\mu V_\mu = \beta
\label{vvst}
\end{equation}

\subsection{A pulse solution in a static background}
Let us now consider a cylindrical source which is static for a
period of time until it starts contracting and emits a sharp pulse
of radiation traveling outward from the axis. Then, the function
$\psi$ can be written as \cite{Carmeli, Herrera}
\begin{equation}
\psi=\frac{1}{2 \pi}\int_{-\infty}^{t-r}
\frac{f(t^{\prime})dt^{\prime}}{[(t-t^\prime)^2-r^2]^{1/2}} + \psi_{LC},
\label{pulse1}
\end{equation}
In (\ref{pulse1}) $\psi_{LC}$ represents the Levi-Civita static solution (\ref{psist}),
and  $f(t)$ is a function of time representing the strength of the source
of the wave and it is assumed to be of
the form
\begin{equation}
f(t)=f_0\delta (t),
\label{pulse2}
\end{equation}
where $f_0$ is a constant and $\delta (t)$ is the Dirac delta function. It
can be shown that (\ref{pulse1}) satisfies the wave equation (\ref{2}).
Then we get
\begin{eqnarray}
\psi=\psi_{LC}, \;\; t<r; \label{pulse3} \\
\psi=\frac{f_0}{2\pi(t^2-r^2)^{1/2}} + \psi_{LC}, \;\; t>r. \label{pulse4}
\end{eqnarray}
from the above, equation (\ref{L2}) can be easily deduced.

The function $\psi$,  as well as its derivatives, is regular
everywhere except at the wave front determined by the surface $t=r$,
followed by a tail decreasing with $t$. 

From (\ref{pulse4}) and   Einstein's equations, we obtain
\begin{equation}
\gamma(r,t) =\beta^2 \ln{r} + \left(\frac{f_0}{\pi}\right)^2 \frac{r^2}{8(t^2-r^2)^2} - \frac{f_0}{\pi} \frac{ \beta}{(t^2-r^2)^{1/2}}
\label{gamp}
\end{equation}
Using (\ref{pulse4}) into (\ref{geo10}) and (\ref{vv1})
we obtain
\begin{equation}
\omega^2 = e^{2\gamma} \left[\frac{\beta}{r^2} - \frac{f_0}{2\pi}\frac{1}{(t^2-r^2)^{3/2}} + \left(\frac{f_0}{\pi}\right)^2 \frac{t^2}{4(t^2-r^2)^3}
\left(1+\beta-\frac{f_0}{2\pi} \frac{r^2}{(t^2-r^2)^{3/2}}\right)^{-1}\right]
\label{omp}
\end{equation}
with $\gamma$ given by (\ref{gamp}).

In the limit $t>>r$, neglecting terms  of order $O(\frac{r}{t})$ and higher we obtain for $\omega^2$
\begin{equation}
\omega^2 = r^{2(\beta^2-1)}\beta e^{-2 \beta f_0/{\pi t}}=\omega_{LC}^2 e^{-2 \beta f_0/{\pi t}}
\label{ompl}
\end{equation}

From the above, it follows that the angular velocity is smaller than in the static case.

Finally, it is worth noticing that just behind the front, in the limit  $t \stackrel{>} \sim r$,
the leading term in (\ref{L}) vanishes identically, supporting the link between
gravitational radiation and the vanishing of $Q$ and $I$ \cite{Bonnora}.

\subsection{The Kramer solution in a static background}
Let us now consider the Kramer solution \cite{Kramer}  in a static background \cite{Stephani}, thus we have.
\begin{eqnarray}
\psi &=& -\beta \ln{r} + C J_0(r) \cos{t}\nonumber \\
\gamma &=& \beta^2 \ln{r} +\frac{1}{2} C^2 r \left\{r \left[J_0^2(r)+J_1^2(r)\right] - 2 J_0(r) J_1(r) {\cos^2{t}}\right\} - 2 C \beta J_0(r) \cos{t}\nonumber\\
\label{kr}
\end{eqnarray}
then from (\ref{geo10}) and (\ref{kr}) we obtain
\begin{eqnarray}
\omega^2 &=& r^{2 \beta^2} \exp{\left[Cr^2 \left\{r(J_0^2(r)+J_1^2(r)) - 2J_0(r)J_1(r)\cos^2{t}\right\}-4C\beta J_0(r) \cos{t}\right]}\nonumber \\
&\times& \left[\frac{\beta}{r^2} + \frac{C J_1(r) \cos{t}}{r} + \frac{C^2 J_0^2(r) \sin^2{t}}{1 + \beta + C r J_1(r) \cos{t}}\right]
\label{omkr}
\end{eqnarray}
and from (\ref{vv1}) and (\ref{kr})
\begin{equation}
V^\mu V_\mu = \beta + CrJ_1 \cos{t} + \frac{r^2 C^2 J_0^2(r) \sin^2{t}}{1 + \beta + C r J_1(r) \cos{t}}.
\label{vvkr}
\end{equation}
The time average  for  (\ref{vvkr}) turns out to be:
\begin{equation}
\left<V^\mu V_\mu\right> = \beta + (1 + \beta) \frac{J_0^2(r)}{J_1^2(r)} \left[1 - \left(1 - \frac{r^2 C^2 J_1^2(r)}{(1 + \beta)^2}\right)^{1/2}\right]
\label{vvkrpr}
\end{equation}

The second term in the right hand side of (\ref{vvkrpr}) being
positive, it is clear that this time average is larger than the
corresponding value in the static case.

\section{Super--Poynting and Lagrangian Poynting  vectors in the Einstein--Rosen spacetime}
The super--Poynting vector as defined in \cite{Basset} is given by
\begin{equation}
P_{\alpha}=\epsilon_{\alpha \beta \gamma \delta}E^{\beta}_{\rho}H^{\gamma \rho}u^{\delta}
\label{p1}
\end{equation}
giving
\begin{equation}
P_2=P_3=0
\label{p2}
\end{equation}
and
\begin{equation}
P_1=\frac{H_{23}e^{\psi-\gamma} \sqrt{-g}}{r^2}(E_{33}\frac{e^{2\psi}}{r^2}-e^{-2\psi}E_{22})
\label{p3}
\end{equation}
or, using (\ref{h23}--\ref{e33})
\begin{eqnarray}
P_1&=& e^{3(\psi-\gamma)}[\psi_{rr}(-6\psi_r \psi_t+2r\psi^3_t+6r\psi_t\psi^2_r)-2\psi_{rr}\psi_{tr} \nonumber \\
&+&\psi_{tr}(-3\psi^2_t-3\psi^2_r-\frac{\psi_r}{r}+6r\psi_r\psi^2_t+2r\psi^3_r)-
8\psi^3_t\psi_r \nonumber\\
&+& 3r\psi^5_t
+30r\psi^3_t\psi^2_{r}-6\psi^3_r\psi_t-\frac{3\psi^2_r \psi_t}{r}+15r\psi^4_r\psi_t\nonumber \\
&-& 6r^2\psi_r\psi^5_t -20r^2\psi^3_t\psi^3_r -6r^2\psi^5_r\psi_t].
\label{p4}
\end{eqnarray}

An alternative expression to evaluate the flux of gravitational energy is given  in \cite{Stephani}. This Poynting vector,
named ``Lagrangian'' by Stephani, yields for the
radial component of the energy flux:
\begin{equation}
S^1=-2r\psi_r \psi_t
\label{lag}
\end{equation}

It can be checked without difficulty that (\ref{lag}) is essentially equivalent to the expressions
obtained from the Landau--Lifshitz \cite{Landau} and Tolman \cite{Tolman}
energy--momentum complex, when calculated in cartesian coordinates (see eqs.(5), (10) and (14) in \cite{Vir}).

For the pulse--type solution, it follows from (\ref{pulse4})
\begin{equation}
S^1=\frac{f_0 t}{\pi (t^2-r^2)^{3/2}}\left[\frac{f_0 r^2}{2\pi (t^2-r^2)^{3/2}}-\beta \right]
\label{s1}
\end{equation}
which  in the limit $t>>r$ (far behind
the pulse) becomes
\begin{equation}
S^1 \approx-\frac{f_0 \beta}{\pi t^2}
\label{s11}
\end{equation}
where we have neglected terms of order $O(\frac{r}{t})$ and higher. This is clearly a negative quantity.

However, just behind the pulse ( $t \approx r, t>r$),
the flux is positive (directed outward) if only
\begin{equation}
\beta<\frac{f_0 r^2}{2 \pi (t^2-r^2)^{3/2}}
\label{c1}
\end{equation}

Next, taking the limit $t>>r$ in the radial component of the
super--Poynting vector we get
\begin{equation}
P^1 \approx e^{3(\psi-\gamma)} \left(- \frac{3 f_0}{2\pi t^2 r^3}\right) \left(\beta^2 + 4\beta^3 + 5 \beta^4 + 2 \beta^5\right)
\label{sptg}
\end{equation}
where we have neglected terms of order $O(\frac{r}{t})$ and higher. This is also a negative quantity.

On the other hand, in the limit $t \approx r (t>r)$ (just behind the pulse) we obtain
\begin{equation}
P^1 \approx \frac{e^{3(\psi-\gamma)}f_0^5 t^5}{2 \pi^5 (t^2-r^2)^{15/2}}\left[\frac{f_0t^2}{\pi(t^2-r^2)^{3/2}}-\frac{7 \beta}{2}\right]
\label{sptr}
\end{equation}
which is positive if
\begin{equation}
\beta<\frac{2f_0 r^2}{7 \pi (t^2-r^2)^{3/2}}.
\label{c2}
\end{equation}

Therefore it is apparent  from the above and (\ref{L2}), that circular geodesics in the pulse--type solution,
exist only when the radial flux of energy (as measured by either the
super--Poynting  or the ``Lagrangian'' vector) is directed inward.

Unfortunately we were unable to calculate the time average of (\ref{p4}) for the Kramer solution, since the resulting integrals cannot be expressed in terms of elementary functions.
From a very rough numerical estimate it seems hoewever, that such average is not vanishing.

In the next section we shall present a discussion on the results obtained so far.
\section{Discussion}
We have seen that E--R waves produce a non--vanishing magnetic Weyl
tensor. This fact together with the vanishing of the vorticity for
the congruence of observers at rest in (\ref{1}) reinforces  the
link between gravitational radiation and the magnetic part of Weyl
tensor. We have seen further, that no purely magnetic E--R waves
exist, as expected for a purely magnetic vacua, with a non--rotating congruence. Also, it was shown that purely electric
solutions of this type are necessarily static. Proving thereby that
a E--R spacetime is a PES if and only if it is static.

Next we have written down the geodesic equations and exhibited the
influence of the gravitational wave on the circular motion of a test
particle, by comparing with the static case.

In the first   example (pulse--type solution) the angular  velocity is smaller than in the static case, this
difference decreases asymptotically as the system gets back to the
static situation. The physical interpretation of this result is
straightforward: we have an initially static system, whose mass per
unit of length is related to $\beta$ \cite{HSG}, next  the source
emits a wave front traveling outwards, carrying out energy.

This explains why  the angular velocity of the particle is
smaller than in the static case. After that, the spacetime behind
the front  tends asymptotically to a static situation with the same
$\beta$ as initially. This explains why the exponent in  (\ref{ompl}) decreases with time. On the other hand it should be
clear that in order to restore the static situation with the same
initial $\beta$ the system should absorb some energy, in order to
compensate for the energy which has been radiated away.

This is indeed the picture which follows from the expressions for
the radial component of the super--Poynting and the Stephani
vectors. Although it is true  that just behind the front, the energy
flux is directed outward, it should be stressed that there are not
circular geodesics there. These are allowed far from the front where terms of the order of $O(\frac{r}{t})$ and higher can be neglected.

In the case of the Kramer solution (in a static background), the time average of the  tangential
velocity of the particle is greater than in the static case. This suggests an
increasing in the active gravitational mass of the source as compared with the static case.
Such an effect might be interpreted in two different ways.

On the one hand, Kramer interprets his solution as describing a
gravitational wave propagating in the $z$-direction. Then the energy
of this wave could account for the increasing of the active
gravitational mass of the source. Observe that the time average of
the radial component of the Stephani vector vanishes in this case,
which is consistent with the interpretation above.

On the other hand, it could be that, as in the pulse case, there is an incoming flow of
energy which is responsible for the increasing of the total energy of the source.  If this is so,
then at least the time average of the radial component of the super--Poynting vector should be negative.

Unfortunately the expression is so complicated that we were unable
to find such average, even with the help of a computer. However, if
this last  interpretation turns out to be correct, then it is clear
that the super--Poynting vector would be  more suitable for
describing the flow of gravitational energy than the ``Lagrangian''
Poynting vector.

Finally, and just as a curiosity, it is worth mentioning that in
classical electrodynamics, the Poynting vector for the field of a
current along an infinite wire (with non--vanishing resistance),
describes a flux of electromagnetic energy directed radially into
the wire \cite{Feymann}.

In  this example, Feynman dismisses the significance of such an
effect by arguing that it is deprived of any physical relevance.

In our case however, the presence of a radial,  inwardly directed flux of gravitational energy  (at least in the example of the pulse
solution), is  necessary in order to restore the energy carried away by the pulse, and, at the same time,
to explain the time evolution pattern of an  ``observable'' quantity such as
the angular velocity of a test particle in a circular motion around the source.

\section*{Acknowledgments.}
One of us (JC) gratefully acknowledges financial support from the
Spanish Ministerio de Educaci\'{o}n y Ciencia through the grant
FPA2004-03666. LH wishes to thank FUNDACION POLAR for financial support and Universitat  Illes Balears for financial support and hospitality. ADP also acknowledges hospitality of the
Physics Department of the  Universitat  Illes Balears.

\end{document}